\documentclass[aps,prx,twocolumn,superscriptaddress,showpacs]{revtex4-1}
\usepackage{amsmath,amssymb,graphics,epsfig,epstopdf,color,verbatim,ulem,braket,tabularx}
\usepackage{multirow}
\usepackage[colorlinks,linkcolor=blue,citecolor=blue,urlcolor=blue,bookmarks=false]{hyperref}

\usepackage{listings}
\usepackage{cancel}
\usepackage{mathrsfs}
\usepackage{soul}
\usepackage{color}
\usepackage{url}

\begin{document}

\title{Doping asymmetry and layer-selective metal-insulator transition in trilayer K$_{3+x}$C$_{60}$}

\author{Changming Yue}
\email{changming.yue@unifr.ch}
\affiliation{Department of Physics, University of Fribourg, 1700 Fribourg, Switzerland}

\author{Yusuke Nomura}
\affiliation{RIKEN Center for Emergent Matter Science, 2-1 Hirosawa, Wako, Saitama 351-0198, Japan}

\author{Philipp Werner}
\email{philipp.werner@unifr.ch}
\affiliation{Department of Physics, University of Fribourg, 1700 Fribourg, Switzerland}

\begin{abstract}
Thin films provide a versatile platform to tune electron correlations and explore new physics in strongly correlated materials. 
Epitaxially grown thin films of the alkali-doped fulleride K$_{3+x}$C$_{60}$, for example, exhibit 
intriguing phenomena, including Mott transitions and superconductivity, depending on dimensionality and doping.
Surprisingly, in the trilayer case, a strong electron-hole doping asymmetry has been observed 
in the superconducting phase, which is absent in the three-dimensional bulk limit. 
Using density-functional theory plus dynamical mean-field theory, we show that this doping asymmetry results from a substantial charge reshuffling from the top layer to the middle layer. 
While the nominal filling per fullerene is close to $n=3$, the top layer rapidly switches to an $n=2$ insulating state upon hole doping, which 
\textcolor{black}{implies a}
doping asymmetry of the superconducting gap. 
The interlayer charge transfer and layer-selective metal-insulator transition
result from the interplay between crystal field splittings, strong Coulomb interactions, and an effectively negative Hund coupling. This peculiar charge reshuffling is absent in the monolayer system, which is an $n=3$ Mott insulator, as expected from the nominal filling. 
\end{abstract}

\maketitle

\newpage

{\it Introduction.} \ 
The alkali-doped fullerides $A_3$C$_{60}$ ($A$ = K, Rb, Cs) exhibit a remarkably high superconducting critical temperature in the range of $20$-$40$ K, and several properties suggest an unconventional pairing mechanism \cite{Prassides2016,Crespi2002,Capone2002,Capone2009,Nomura2016}.
The materials are strongly correlated three-orbital systems, with an intra-orbital interaction $U$ comparable to the bandwidth $W$ of about 0.5 eV \cite{Nomura2012}. 
Intriguingly, $s$-wave superconductivity, which had been believed to be fragile to strong electron correlations, appears in the vicinity of a Mott phase \cite{Prassides2016}. 
As $U/W$ is tuned via chemical or physical pressure, a $T_c$ dome is observed \cite{Zadik2015}, and the metallic phase above this dome exhibits unusual properties on the strong-coupling side. More specifically, a Jahn-Teller metal with coexisting metallic and Mott insulating orbitals has been experimentally observed \cite{Zadik2015} and theoretically explained as a spontaneous orbital selective Mott phase \cite{Hoshino2017}. The local singlet pairing and unusual normal-state properties of fulleride superconductors originate from the negative effective Hund coupling $J\approx -0.02$ eV, which results from the overscreening of the very small static  
Hund coupling $J_\text{Coulomb}\approx 0.03$ eV in this molecular crystal by Jahn-Teller phonons \cite{Nomura2015}, $\Delta J_\text{Jahn-Teller}\approx -0.05$ eV. This produces an orbital-freezing in the strongly correlated metal regime, and the associated local orbital fluctuations have been suggested 
to play a key role in the pairing mechanism \cite{Hoshino2017,Yue2021}, in contrast to the spin-freezing that is commonly associated with  unconventional superconductivity in positive-$J$ systems \cite{Hoshino2015,Werner2016}. While many interesting properties and (doping-dependent) phase diagrams for $A_3$C$_{60}$ type systems have been revealed by model and \textit{ab initio} calculations \cite{Crespi2002,Capone2002,Nomura2015,Hoshino2017,Yue2020,Yue2021}, experimental progress has been hampered by difficulties in preparing bulk single crystals. 

Thin films offer a promising route to explore an expanded phase space of correlated materials. 
Many strongly correlated systems exhibit novel properties at interfaces or in the limit of few atomic layers. Prominent examples are correlated metallic or  superconducting states at the interface between Mott and band insulators \cite{OkamotoMillis2004,Science2007_317_1196}, 
or the remarkably high superconducting $T_c$ in monolayer FeSe \cite{CPL2012_SC_FeSe, NatCom2012_SC_FeSe, He_NatMat_2013_SC_FeSe, Tan_NatMat_2013_SC_FeSe, CPL2014, PRL2016_SC_FeSe, AnuRev2017_FeSe}. It is thus an interesting question what kind of correlation effects appear in fulleride compounds as one reduces the dimensionality from 3D to 2D.  

\begin{figure}
\includegraphics[clip,width=2.6in,angle=0]{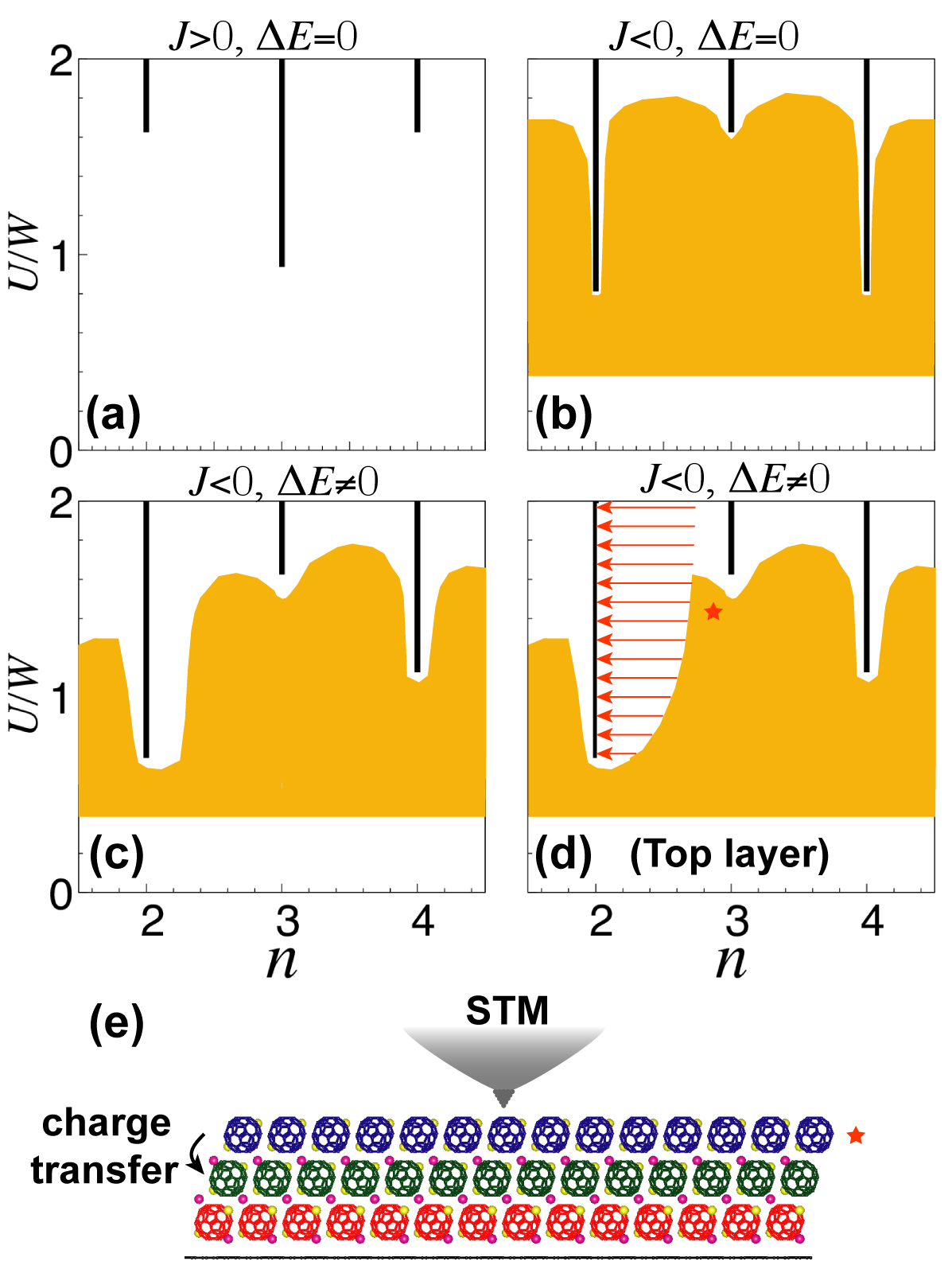}
\caption{
Schematic phase diagram of three-orbital systems. (a) Generic phase diagram of an orbitally degenerate three-orbital Hubbard model with $J>0$. The black lines indicate the Mott insulating solutions with filling $n=2$, $3$, and $4$ and the white region a metallic solution. (b) Generic phase diagram of an orbitally degenerate three-orbital Hubbard model with $J<0$. 
The yellow area shows the stability region of the $s$-wave superconducting phase at low temperatures. (c) Modification of the phase diagram by a crystal field splitting $\Delta E$ of the type ``two-up, one down," which favors the $n=2$ insulating state.
(d) Surface layer of the trilayer system. The red arrows indicate that there is no solution in the corresponding filling region, because of interlayer charge transfer which results in $n=2$. The red star indicates the filling in the nominally undoped trilayer system. 
(e) STM measures the properties of the top layer. 
}
\label{fig:illustration}
\end{figure}

Thin films of $A_3$C$_{60}$ have been realized already more than 20 years ago \cite{Geerligs2000,ZXShen2003_MLK3C60},
but the recent systematic study of the superconducting properties of 
 epitaxially grown high-quality thin films of K$_{3+x}$C$_{60}$ 
 represents an important step towards a detailed investigation of fulleride compounds \cite{Xue_prl_2DK3C60_SC}. These multi-layer systems are single crystals, which allow to study the effects of dimensionality (3D $\rightarrow$ 2D) and number of layers. Scanning tunneling microscopy (STM) can be used to obtain accurate information on the electronic states in the top layer \textcolor{black}{(Fig.~\ref{fig:illustration}(e))}. Furthermore, by varying the concentration of the alkali atoms, these three-orbital systems can be doped over a wide range $n=3+x$ relative to the half-filled stoichiometric compound ($n=3$).  The STM results \cite{Xue_prl_2DK3C60_SC} demonstrated a Mott insulator to metal (or superconductor) transition with increasing number of layers, which may be explained by the larger screening and connectivity in the 3D limit. More surprisingly, in the superconducting trilayer samples, a very strong asymmetry in the gap size with respect to electron and hole doping was observed. While hole doping quickly destroys the superconducting state, the latter is remarkably robust to electron doping. This is in stark contrast to 
 3D bulk systems \cite{Yildirim1996}, 
 which do not show any significant doping asymmetry. Filling-dependent changes in screening properties have been suggested as a possible mechanism \cite{Xue_prl_2DK3C60_SC}, but this is unlikely to fully explain the observed strong asymmetry.

Here, using realistic simulations of trilayer K$_{3+x}$C$_{60}$ based on density functional theory (DFT) \cite{Kohn1965} and dynamical mean-field theory (DMFT) \cite{Georges1996}, we show that charge reshuffling between the layers and a layer-selective metal-insulator transition is at the origin of the doping asymmetry. In particular, even though the naively expected filling is close to $n=3$, hole doping quickly leads to the formation of an $n=2$ insulating state in the top layer, with one almost fully occupied and two almost empty orbitals, while electron doping results in three partially filled metallic orbitals. 

{\it Metal-insulator transitions in the negative-$J$ model} \ 
To set the stage, we first discuss the generic phase diagram of the negative-$J$ three-orbital 
Hubbard model (Fig.~\ref{fig:illustration}), considering overscreening of $J$ by phonons~\cite{Capone2009,Nomura2016}.
In contrast to positive-$J$ multi-orbital systems, where the Hund coupling stabilizes the half-filled $n=3$ Mott state, relative to the neighboring $n=2$ and $n=4$ Mott states [panel (a)], the effective $J<0$ in K$_{3+x}$C$_{60}$ destabilizes the $n=3$ Mott phase and pushes the critical on-site interaction to higher values [panel (b)]. In the absence of crystal-field splittings or asymmetries in the density of states (DOS), 
the $n=2$ and $n=4$ Mott states are equivalent (particle-hole symmetry), but in a layered system, a ``two up, one down" splitting is introduced (see 
Tab.~\ref{tab:Eeff_U0p7}), which favors the $n=2$ insulator state with one almost completely filled and two almost empty orbitals [panel (c)]. This is a correlation-induced orbitally-polarized insulator \cite{Tosatti1997,Tosatti2000}. 
The insulating regions in Fig.~\ref{fig:illustration} were obtained for a semi-circular density of states, but realistic parameters and energy splittings. 

The boundary of the $s$-wave superconducting phase is sketched in yellow based on the results reported in Ref.~\onlinecite{Yue2020}.  Considering the difference in dimensionality, 
one can imagine a situation where the half-filled ($n=3$) monolayer is Mott insulating, and doping it does not result in a superconducting state (consistent with Ref.~\onlinecite{Wu2006}), 
while the half-filled trilayer system is superconducting. 
In the trilayer case, because of the asymmetric situation, each layer can have a different filling and correlation strength. 
When a strongly correlated layer has a filling of $n<3$, the existence of a very stable $n=2$ insulating state may trigger a drastic charge reshuffling associated with a layer-selective metal-insulator transition. 
The realistic simulations below will show that this scenario is indeed realized in the top layer: 
the filling of the top layer of the K$_{3+x}$C$_{60}$ thin film is below half-filling already at $x=0$ (see red star in panel (d)), and the top layer quickly switches into an $n=2$ insulating state upon hole doping $x<0$ (red arrows in panel (d)) \textcolor{black}{and charge transfer to the middle layer (black arrow in panel (e)).}

\begin{figure*}[htp]
\includegraphics[clip,width=0.8\paperwidth,angle=0]{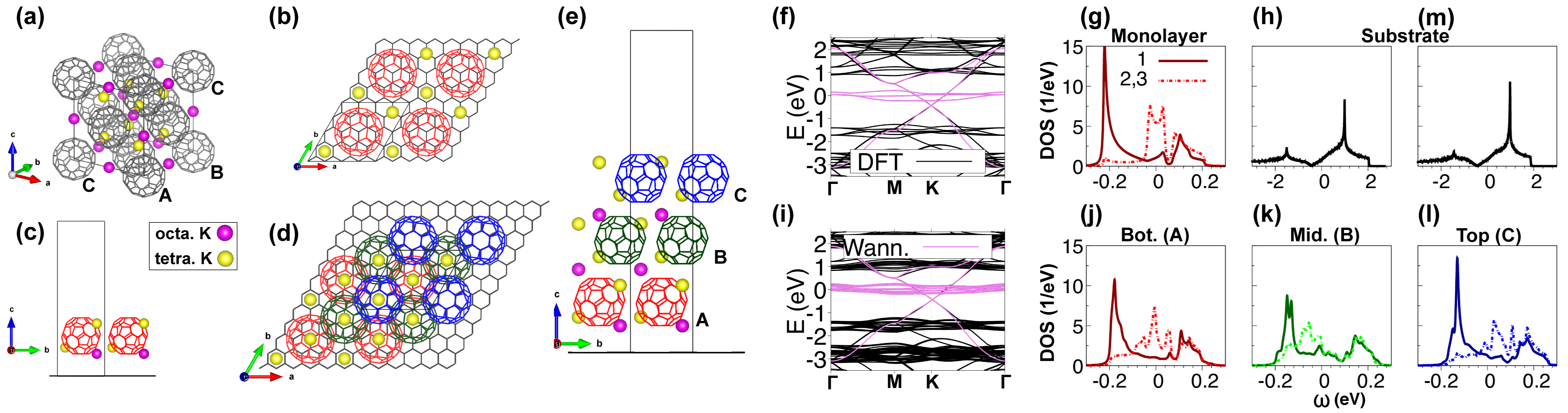}
\caption{ Crystal structure and electronic structure of monolayer and trilayer K$_3$C$_{60}$ with graphene substrate.
(a) Conventional unit cell of bulk K$_3$C$_{60}$. 
(b-c) [(d-e)] Top and side view of monolayer [trilayer] K$_3$C$_{60}$ with graphene substrate (honeycomb lattice).
For better visualization, the bottom (A), middle (B) and top layer (C) of C$_{60}$ in (b-e)
are colored red, dark-green and blue, respectively. 
The primitive unit cell is marked by the rhombus in (b) and (d).
The DFT (black) bands and their Wannier interpolations (violet) are shown in panel (f) for the monolayer and in panel (i) for the trilayer, respectively. 
The corresponding orbital resolved  
DOS per spin for each layer are shown in panel (g) and (j-l), 
where the solid line with dark color shows the projected DOS for the lowest-energy orbital and
the dot-dashed line with lighter color that for the degenerate higher-energy orbitals. Panels (h) and (m) plot the total DOS for the substrate in (c) and (d), respectively. 
}
\label{fig:struct}
\end{figure*}

{\it Ab initio calculations} \ 
Bulk K$_3$C$_{60}$ has a face-centered cubic (FCC) structure, which can be viewed as close packing of (111) planes in a periodic sequence of ABC ABC layers along the [111] direction, as shown in Fig.~\ref{fig:struct}(a). 
The K$_{3+x}$C$_{60}$ films with small doping $x$  also crystallize into an FCC structure \cite{Xue_prl_2DK3C60_SC}. 
Monolayer K$_3$C$_{60}$ consists of only the A layer [panels (b),(c)], while the trilayer system is constructed by stacking the A, B, and C layers [panels (d),(e)].
To study the effect of the substrate, we place the films on top of a single layer of graphene. Details of the DFT calculations and structure relaxation are provided in the Supplemental Material (SM).
The resulting DFT band structures are shown in Fig.~\ref{fig:struct}(f), and Fig.~\ref{fig:struct}(i). In the case of the monolayer (trilayer) + substrate system, there are 3 (9) narrow $t_{1u}$ bands derived from C$_{60}$ and three wide bands from the graphene substrate near the Fermi energy. The tight binding Hamiltonian is obtained from the maximally-localized Wannier functions constructed by wannier90 \cite{wannier90,Pizzi_2020}, and reproduces the DFT dispersion near the Fermi energy, see violet lines in panels (f) and (i). 

We consider the orbitals which diagonalize the onsite Hamiltonian. 
For the trilayer system, the energy splittings between these orbitals are listed in 
Tab.~\ref{tab:Eeff_U0p7}.
The splittings are substantial 
 in the case of the monolayer (118 meV) and the top layer of the trilayer system (82.7 meV). The corresponding orbital-resolved DOS is shown in Fig.~\ref{fig:struct}(g) for the monolayer and in Fig.~\ref{fig:struct}(j),(k),(l) for the trilayer. On the other hand, the asymmetry in the total DOS is not sufficiently strong near the Fermi level that the DFT picture could explain the experimentally observed doping asymmetry. We also show, in panels (h) and (m), the DOS of the substrate in the monolayer [(h)] and trilayer [(m)] system, which due to the strong dispersion of the bands is small near the Fermi energy.  

The DFT+DMFT calculations are performed using maximally localized Wannier functions \cite{PRB_WannDMFT2006} without charge self-consistency. 
Each layer is mapped to a three-orbital Anderson impurity model embedded in a self-consistent electron bath \cite{Nolting_1999,layerDMFT_PRB2015}.
The self-energy for each layer is assumed to be local and orbital-diagonal, and the DMFT impurity problems are solved with local density-density interactions 
($H_{\mathrm{int}}=\sum_{\alpha} U n_{\alpha \uparrow} n_{\alpha \downarrow}\quad+\sum_{\sigma, \alpha>\gamma}[(U-2 J) n_{\alpha \sigma} n_{\gamma \bar{\sigma}}+(U-3 J) n_{\alpha \sigma} n_{\gamma \sigma}]$ with $n_{\alpha\sigma}$ the spin-density in orbital $\alpha$) using continuous-time quantum Monte Carlo simulations in the hybridization expansion (CT-HYB) implementation \cite{Werner2006,Gull2011}.
 We choose realistic parameters for the on-site interaction ($U=0.7$ eV) and effective Hund coupling ($J=-0.03$ eV), \textcolor{black}{consistent with a recent {\it ab initio} study \cite{Nomura2015} and constrained random phase approximation estimates \cite{Nomura2012,Nomura2015b}, and present results for temperature $T=50$ K. 
Additional simulation details, including the results for a rotationally invariant interaction and the double counting correction used to shift the weakly correlated substrate bands relative to the strongly correlated fulleride bands, are discussed in SM, where we also demonstrate that our main conclusions do not rely on a fine-tuning of parameters.}

\begin{figure}[htp]
\includegraphics[clip,width=0.4\paperwidth,angle=0]{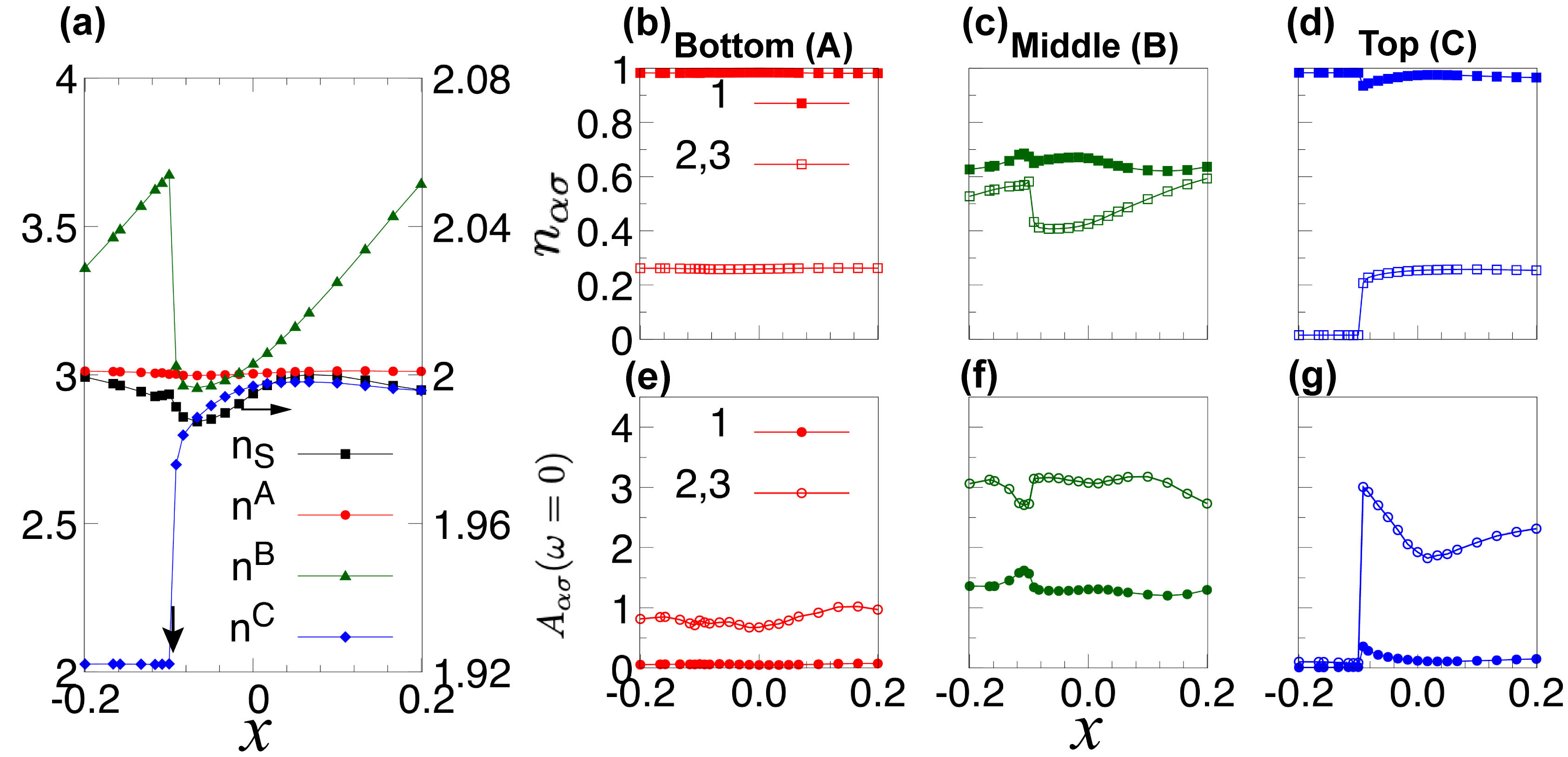}
\caption{DFT+DMFT results for the trilayer ${\rm K}_{3+x}{\rm C}_{60}$+substrate system at $T=50$ K, $U=0.7$ eV, $J=-0.03$ eV. 
(a) Fillings in the different layers and in the substrate (S) as a function 
of doping $x$.
(b-d) Orbital-resolved fillings per spin $\sigma$ in the different layers as a function of $x$.
(e-g) Orbital-resolved spectral weight per spin at $\omega=0$ in the different layers as a function of $x$.
}
\label{fig:trilayer_U0p7}
\end{figure}

{\it Results} \ 
The filling $n^\text{A,B,C}$ in the different layers is plotted as a function of doping $x$ in Fig.~\ref{fig:trilayer_U0p7}(a). 
The first noteworthy observation is that relative to the nominal value of $n=3$ per layer, the top layer (C) is slightly hole doped ($n^\text{C}\approx 2.96$), and layer B slightly electron doped ($n^\text{B}\approx 3.04$), 
layer A undoped ($n^\text{A}\approx 3.00$), while there is little charge transfer from/to the substrate (black lines in panel (a); note the different scale). 
There is hence a charge reshuffling from the top layer with the lowest connectivity to the middle layer with the highest connectivity already in the half-filled system.  
Upon hole doping, we observe a significant additional charge transfer from the top layer C to the middle layer B \textcolor{black}{(Fig.~\ref{fig:illustration}(e))}, and the switching of the top layer C into a state with filling $n^\text{C}\approx 2$. Besides a sheet of potassium atoms in the octahedral position, the presence of the substrate introduces an asymmetry between the bottom and the top layer. As a result of strong correlations, crystal field splittings, and this asymmetry, the trilayer system undergoes a layer-selective metal-insulator transition at small hole doping, while on the electron-doped side, a qualitatively different behavior with a stable metallic solution is found over a wide doping range. 

More detailed information on the evolution of the doping in the different orbitals, as well as the spectral weight at the Fermi level, $A_{\alpha\sigma}(\omega=0)$, is shown in panels (b)-(d) and (e)-(g), respectively. The results for layer A show that the occupation and low-energy spectral weight in this layer are only weakly dependent on doping $x$. The lower orbital 1 is essentially full and hence band-insulating, while the two higher-lying orbitals 2,3 are roughly quarter-filled and metallic (see value of the spectral weight), but the almost vanishing charge compressibility indicates that they are on the verge of becoming Mott insulating. In the top layer C, all three orbitals undergo a metal-insulator transition around $x=-0.092$. Below this doping, the layer is in an orbitally-polarized insulating state with an almost completely filled orbital 1 and almost empty orbitals 2,3. Above this doping, orbitals 2 and 3 are metallic with a filling and spectral weight that depends only weakly on $x$. As demonstrated in Ref.~\onlinecite{Werner2007} (for a positive-$J$ system), the transition from metal to orbitally polarized insulator depends sensitively on the details of the parameters $U$, $J$ and the energy splitting $\Delta E = E_{2,3}-E_{1}$. In particular, the effective level splittings can be substantially enhanced by the real part of the self-energy $\Sigma_\alpha(\omega=\infty)$, which is also the case here, as one can see by comparing the effective level splittings $\Delta \tilde E$ (including the real part of the self-energy) in Tab.~\ref{tab:Eeff_U0p7} to the bare splittings $\Delta E$  
\textcolor{black}{shown in brackets}. 
The splittings in layers C and A are strongly enhanced, and the transition to the orbitally-polarized insulator is associated with a further increase in the effective splitting in layer C. A qualitative difference to the previously discussed $J>0$ model is that in the $J<0$ system, the Hund coupling stabilizes the orbitally-polarized insulator, since the intra-orbital interaction is smaller than the inter-orbital one.  

The orbitals in the most weakly correlated middle layer B remain metallic for all measured dopings $x$. By comparison between panels (c) and (d) we conclude that the metal-insulator transition in the top layer is associated with a transfer of charge to the higher-lying orbitals 2,3 in the middle layer. 
Initially these orbitals absorb the holes, which results in an increasing orbital polarization with decreasing $x$, until the charge transfer from layer $C$ associated with the metal-insulator transition reduces it substantially. Also upon electron doping, the extra charges are essentially absorbed by orbitals 2,3 in the middle layer. 

\begin{table}
\centering
\caption{ The effective level splittings $\Delta\tilde{E}=\tilde{E}_{2,3}- \tilde{E}_{1}$ of the trilayer K$_{3+x}$C$_{60}$+substrate system 
for different dopings $x$ at $T=50$ K, $U=0.7$ eV, and $J=-0.03$ eV. Here $\tilde{E}_i=E_{i}+\mathrm{Re}\Sigma_{i}^{\infty}$ is the effective energy level for the $i$-th orbital in a given layer. The numbers in parentheses are the bare crystal field splittings $\Delta E=E_{2,3}-E_1$.
The energy unit is meV. 
}
\label{tab:Eeff_U0p7}
\medskip
\begin{tabular}{@{\  }c@{\ \ \ }c@{\ \ \ }c@{\ \ \ }c@{\  }}
Doping $x$ &      Bot. (A) & Mid. (B) & Top (C)\tabularnewline
\hline
$-$0.1  &  679.2  & 114.5 & 908.1  \tabularnewline
  0   &  681.6 (61.8)  & 230.0 (21.1) & 696.9 (82.7)  \tabularnewline
 0.1  &  677.7 & 113.5  & 691.0  \tabularnewline
\hline
\end{tabular}
\end{table}

\begin{figure}
\includegraphics[clip,width=3.4in,angle=0]{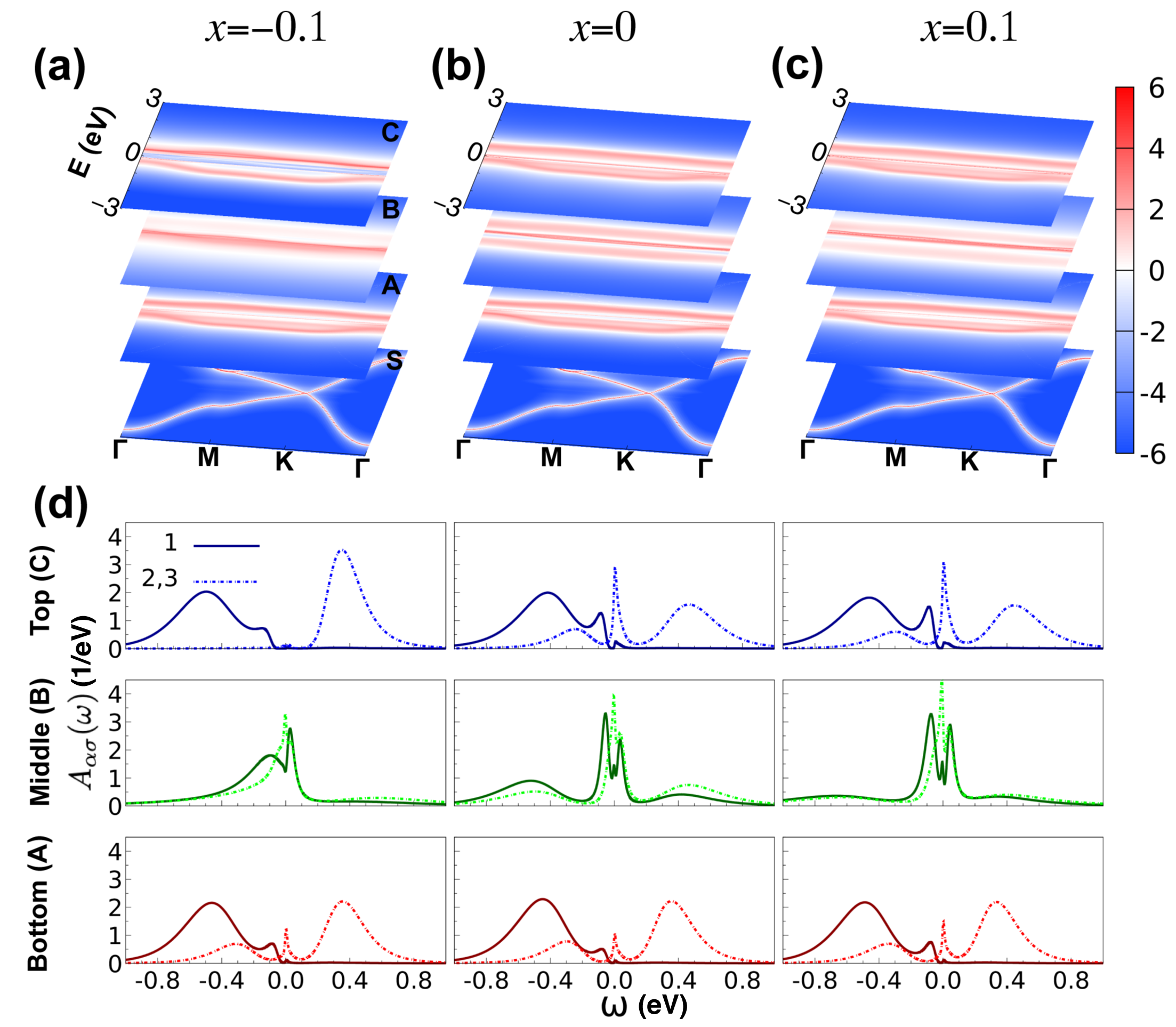}
\caption{DFT+DMFT spectra for the trilayer ${\rm K}_{3+x}{\rm C}_{60}$+substrate system at $T=50$ K, $U=0.7$ eV, $J=-0.03$ eV.
(a-c) Momentum-resolved spectra $\log A(\bf{k},\omega)$ for doping $x=-0.1$, $0$, and $0.1$. 
A, B, C label the different layers and S the substrate.
(d) Corresponding orbital-resolved 
local spectra per spin $A_{\alpha\sigma}(\omega)$. The thick lines correspond to orbital $1$ and the dashed lines to orbitals $2,3$.
}
\label{fig:trilayer_akw_U0p7}
\end{figure}

The layer-selective metal-insulator transition is also evident in the momentum-resolved spectral functions plotted in Fig.~\ref{fig:trilayer_akw_U0p7}. Panel (b) shows the spectra of all layers and the substrate for the undoped system with $x=0$, while panel (a) [(c)] shows the analogous results for a hole (electron) doped system with $x=-0.1$ ($x=0.1$). The local 
spectra for the two types of orbitals are illustrated for layers A, B, C in the corresponding subpanels of panel (d). The peaks near the Fermi level in the almost empty orbitals 2,3 of the top layer result from hybridization with the partially filled orbital 1 \cite{Yue2020,Pruschke_2002}.

{\it Discussion and Conclusions} \ 
An orbitally-polarized insulator will not become superconducting at even lower temperatures~\cite{Xue_prl_2DK3C60_SC},  
while the metallic state in the considered parameter regime can be expected to do so \cite{Yue2020}. Our finding of a strong doping asymmetry and metal-insulator transition in the normal state thus provides a natural explanation for the experimentally observed asymmetry of the superconducting gap in the STM investigation of the trilayer structure. 
While the $U$ value employed in this study (0.7 eV) is consistent with the {\it ab initio} study in Ref.~\cite{Nomura2015},
we have confirmed our scenario for the asymmetry is robust with respect to a 10--20 \% change in $U$ (see SM).
Also, the spectrum for the monolayer ($n=3$ Mott insulator) is consistent with the STM spectrum~\cite{Xue_prl_2DK3C60_SC}. 
The layer dependence of $U$ may play some role in the different properties of the monolayer and trilayer systems, as discussed in Ref.~\cite{Xue_prl_2DK3C60_SC}, but 
our results suggest that the main difference arises from the interlayer hoppings, which favors metallicity and enable the charge reshuffling associated with the layer-selective metal-insulator transition in the trilayer system.

In summary, the strong doping asymmetry observed in trilayer K$_{3+x}$C$_{60}$ is a consequence of correlation-enhanced crystal field splittings in few-layer thin films and the unusual properties of three-orbital systems with $J<0$. The negative $J$ stabilizes the $n=2$ insulating state relative to the half-filled Mott state, so that hole doping in combination with inter-layer charge transfer results in a metal-insulator transition in the surface layer for weak hole doping. 
The experimental results of Ref.~\onlinecite{Xue_prl_2DK3C60_SC} are thus an interesting manifestation of the effectively inverted Hund coupling in fulleride compounds. 
Interesting open questions concern the effect of charge self-consistency, which however will be numerically very challenging for the trilayer system with 221 atoms per unit cell, the quantitative effect of the pair hopping term, which further enhances the inter-layer charge transfer (see SM), and the study of the superconducting phase at even lower temperatures.

{\it Acknowledgements. ---}
We are grateful to Ming-Qiang Ren for valuable discussions on the experiments reported in Ref.~\onlinecite{Xue_prl_2DK3C60_SC}. The DMFT calculations were performed on the Beo05 cluster at the University of Fribourg, using a code based on iQIST \cite{HUANG2015140,iqist}. C.Y. and P.W. acknowledge support from SNSF Grant No. 200021-196966. 
Y.N. was supported by Grant-in-Aids for Scientific Research (JSPS KAKENHI) (Grant No. 20K14423, 21H01041) and MEXT as ``Program for Promoting Researches on the Supercomputer Fugaku'' (Basic Science for Emergence and Functionality in Quantum Matter ---Innovative Strongly-Correlated Electron Science by Integration of ``Fugaku'' and Frontier Experiments---) (Grant No. JPMXP1020200104).

\begin{widetext}
\begin{flushleft}

\section*{SUPPLEMENTARY MATERIAL}

\subsection*{SM1. Methods}

\subsubsection*{Relaxation} 

The Vienna ab initio simulation package (VASP) \cite{vasp_ref1,vasp_ref2,vasp_ref3} is used for the density functional theory (DFT) calculations and relaxation of the structures. 
The film structures are relaxed under the constraint of $C_3$ symmetry. We fix the vacuum thickness as large as 20 \AA \ (by fixing the length of vector $\boldsymbol{c}$) for all systems. 
We choose a set of lengths for the vector $|a| \equiv|\boldsymbol{a}|$ (=$|\boldsymbol{b}|$) and 
relax only the ion positions for a certain value of $|a|$ without relaxing the cell shape and cell volume (thus enforcing the $C_3$ symmetry).
The ionic relaxation is stopped when the change in the total energy between steps becomes smaller than 2 meV. 
It turns out that the resulting optimal vector length $|a|$ for the trilayer (monolayer) is  9.970 (9.917) \AA, which is  very close to the experimental value of $10.0\pm 0.1$ \AA \, \cite{Xue_prl_2DK3C60_SC}.

\subsubsection*{DFT+DMFT for Inhomogeneous Systems}  

Here we present the details of our DFT+DMFT implementation for inhomogeneous
system, which follows Ref.~\onlinecite{Nolting_1999}. The inhomogeneous system
consists of the K$_3$C$_{60}$ multilayer and the substrate. The basic idea is to assume
that each layer has a local self-energy,  i.e. $\Sigma_{AB}=\delta_{AB}\Sigma_{A}$, with $A,B$ denoting the layer index.
The following equations are written for the trilayer+substrate
system, but can be straightforwardly extended to systems with an arbitrary
number of layers. 
The Wannier Hamiltonian for the inhomogeneous system is obtained by Wannier90 \cite{wannier90,Pizzi_2020} interfaced to VASP. 
To reduce the hybridization between the original Wannier orbitals, we diagonalize the on-site part of the Wannier Hamiltonian and switch to the so-called crystal-field basis or natural orbitals \cite{PRB_WannDMFT2006}.
The tight-binding Hamiltonian in this basis is denoted as ${\mathcal H}({\bf k})$ and its diagonal onsite part is denoted as $E_{\mathrm{loc}}^{i}$ [$=\frac{1}{N}\sum_{{\bf k}} {\mathcal H} ({\bf k})$]. 
As in the main text, we label the substrate and three layers by $S$, $A$, $B$ and $C$. The lattice Green's function reads
\begin{equation}
G_{\mathrm{lat}}({\bf k},i\omega_{n})^{-1}=(i\omega_{n}+\mu)\mathbb{I}_{24}-{\mathcal H} ({\bf k})-\left[\begin{array}{cccc}
\Sigma_{\mathrm{loc}}^{A}(i\omega_{n})-\Sigma_{\mathrm{dc}}^{A}\mathbb{I}_{6}\\
 & \Sigma_{\mathrm{loc}}^{B}(i\omega_{n})-\Sigma_{\mathrm{dc}}^{B}\mathbb{I}_{6}\\
 &  & \Sigma_{\mathrm{loc}}^{C}(i\omega_{n})-\Sigma_{\mathrm{dc}}^{C}\mathbb{I}_{6}\\
 &  &  & \mu_{S}\mathbb{I}_{6}
\end{array}\right],
\label{eq:DMFT_eq_1}
\end{equation}
where each $\Sigma$-block is a diagonal 6$\times$6 matrix in spin/orbital space.
Since we treat the fulleride films in DMFT and the substrate (graphene) at the DFT level, a double-counting term 
$\Sigma_{\mathrm{dc}}^{i}$ must be introduced to properly align the energy levels of the two subsystems. We adopt here the fully-localized limit double counting scheme \cite{PRB_DC_FLL1993}, but add an additional constant upward shift $\mu_S=0.7$ eV to the substrate's onsite energy.
We discuss in SM.4 the reason why $\mu_S$ is necessary and explain how it is determined.

The local Green's function is obtained by momentum averaging 
\begin{equation}
G_{\mathrm{loc}}(i\omega_{n})=\frac{1}{N}\sum_{{\bf k}}G_{\mathrm{lat}}({\bf k},i\omega_{n}).
\label{eq:DMFT_eq_2}
\end{equation}
According to the self-consistency condition of DMFT, $G_{\mathrm{imp}}(i\omega_{n})=G_{\mathrm{loc}}(i\omega_{n})$, where 
\begin{equation}
\left[G_{\mathrm{imp}}(i\omega_{n})\right]^{-1}=(i\omega_{n}+\mu)\mathbb{I}_{24}-E_{\mathrm{loc}}-\Delta_{\mathrm{imp}}(i\omega_{n})-\left[\begin{array}{cccc}
\Sigma_{\mathrm{imp}}^{A}(i\omega_{n})-\Sigma_{\mathrm{dc}}^{A}\mathbb{I}_{6}\\
 & \Sigma_{\mathrm{imp}}^{B}(i\omega_{n})-\Sigma_{\mathrm{dc}}^{B}\mathbb{I}_{6}\\
 &  & \Sigma_{\mathrm{imp}}^{C}(i\omega_{n})-\Sigma_{\mathrm{dc}}^{C}\mathbb{I}_{6}\\
 &  &  & \mu_{S}\mathbb{I}_{6}
\end{array}\right].
\label{eq:DMFT_eq_3}
\end{equation}
Assuming that $\Sigma^i_\text{loc}=\Sigma^i_\text{imp}$ (DMFT approximation), we thus obtain
the hybridization function $\Delta^i_\text{imp}$ of the Anderson impurity model for layer $i=A,B,C$, which reads 
\begin{equation}
\Delta_{\mathrm{imp}}^{i}(i\omega_{n})=(i\omega_{n}+\mu)\mathbb{I}_{6}-E_{\mathrm{loc}}^{i}-\left[G_{\mathrm{loc}}(i\omega_{n})\right]_{ii}^{-1}-\left[\Sigma^{i}_\text{imp}(i\omega_{n})-\Sigma_{\mathrm{dc}}^{i}\mathbb{I}_{6}\right].
\label{eq:DMFT_eq_4}
\end{equation}
We separately solve the three Anderson impurity models with the hybridizations $\Delta_{\mathrm{imp}}^{i}(i\omega_{n})$ and effective local energy levels $E_{\mathrm{loc}}^{i}-\Sigma_{\mathrm{dc}}^{i}\mathbb{I}_{6}$ using the hybridization-expansion continuous-time quantum Monte-Carlo method (CT-HYB) \cite{Werner06}. 
This gives the new impurity Green's functions $G_{\mathrm{imp}}^{i}$ and hence the new impurity self-energies 
\begin{equation}
\Sigma_{\mathrm{imp}}^{i}(i\omega_{n})=(i\omega_{n}+\mu)\mathbb{I}_{6}-(E_{\mathrm{loc}}^{i}-\Sigma_{\mathrm{dc}}^{i}\mathbb{I}_{6})-\Delta_{\mathrm{imp}}^{i}(i\omega_{n})-\left[ G_{\mathrm{imp}}(i\omega_{n})\right]_{ii}^{-1}.
\label{eq:DMFT_eq_5}
\end{equation}
Then we update the hybridization functions for the next iteration
by plugging $\Sigma_{\mathrm{loc}}^{i}=\Sigma_{\mathrm{imp}}^{i}$ into Eqs.~(\ref{eq:DMFT_eq_1}), (2) and (\ref{eq:DMFT_eq_4}), and then solve the impurity model until convergence.
The symmetric improved estimators \cite{Kaufmann2019} are used in CT-HYB to strongly reduce the noise level in the impurity self-energy, which is very helpful to stabilize the DFT+DMFT iterations.

\subsection*{SM2. DFT+DMFT results for the monolayer+substrate system} 

Figure~\ref{fig:monolayer_U0p7} shows the DFT+DMFT results for the monlayer+substrate system at $U=0.7$ eV, $J=-0.03$ eV, $T=50$ K.
The momentum-resolved spectral functions are shown in Fig.~\ref{fig:monolayer_U0p7}(a), 
and the corresponding orbital-resolved local spectra are shown in Fig.~\ref{fig:monolayer_U0p7}(b), indicating a Mott insulating monolayer system. The monolayer is half-filled, with the low-lying orbital 1 almost completely full ($n_{1\sigma}\approx 1$) and the two degenerate high-lying orbitals 2, 3 quarter filled ($n_{2,3\sigma}\approx 1/4$).

\begin{figure*}[htp]
\centerline{\epsfig{figure=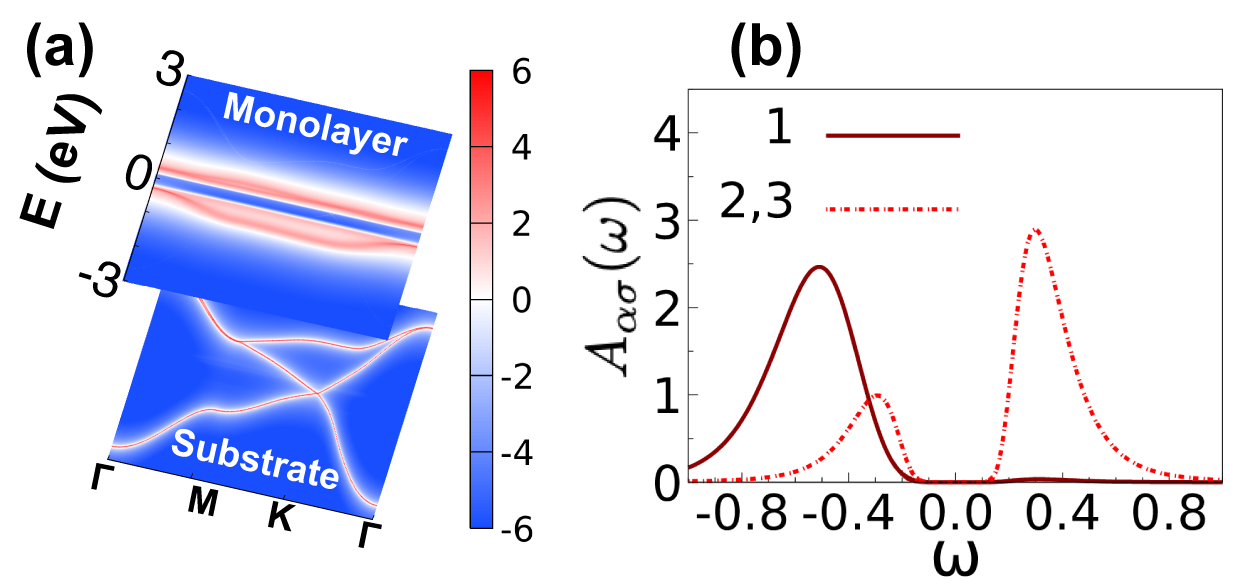,width=0.5\columnwidth}}
\caption{DFT+DMFT spectra for the monolayer ${\rm K}_3{\rm C}_{60}$+substrate system at $T=50$ K, $U=0.7$ eV, $J=-0.03$ eV.
(a) Momentum-resolved spectra $\log A(\bf{k},\omega)$.
(b) Corresponding orbital resolved local spectra per spin A$_{\alpha\sigma}(\omega)$. The thick line corresponds to orbital 1 and the dashed line to orbitals 2, 3. 
}\label{fig:monolayer_U0p7}
\end{figure*}

\subsection*{SM3. DFT+DMFT results for the trilayer+substrate system at different $U$}

Ref. \onlinecite{Xue_prl_2DK3C60_SC} discussed the layer dependence of $U$, and suggested $U\approx 0.2$ eV for the trilayer sytem, which is significantly smaller than the estimate for the monolayer ($U\approx 0.6$ eV).
In the experimental paper, the peaks closest to the gap edge were used to estimate $U$, but from a theoretical point of view, one would rather expect that the center positions of the broad bands determine $U$.
Thus, while there should be some layer-dependence of $U$, it is unclear how reliable the experimental estimates are. 
In this study, instead, we use the {\it ab initio} derived value of $U$ for the bulk system \cite{Nomura2015}. 

According to previous theoretical studies on the filling and layer-thickness dependence of the screening effect, the change in $U$ between bulk systems and few layer systems is on the order of 10-20\%. For example, 
in a system consisting of SrVO$_3$ layers on a SrTiO$_3$ substrate \cite{Zhong_2015}, it was shown that the constrained random phase approximation (cRPA) $U$ for monolayer (bilayer) SrVO$_3$ is only about 20\% (10\%) larger than for the bulk material. Other theoretical studies \cite{Nomura2012_PRB,Han2021_PRB}
suggest that the filling-dependence of $U$ should also be of the order of 10\%.
It is useful to confirm that our results are robust against such variations in $U$.

Fig.~\ref{fig:U0px_Jm0p03_muS_0p7} illustrates the $U$ dependence of the DFT+DMFT results around $U=0.7$ eV. 
Qualitatively, the results for $U=0.6$--$0.8$ eV are all consistent with our interpretation of the experiments in terms of a layer-selective metal-insulator transition, which shows that our main result does not depend on a fine-tuning of the parameter $U$. 
At the quantitative level, the doping value for the layer-selective metal-insulator transition changes slightly. 
Compared with $U=0.7$, a weaker $U=0.6$ (stronger $U=0.8$) shifts the Mott transition point slightly to the electron (hole) doped side, 
see Fig.~\ref{fig:U0px_Jm0p03_muS_0p7}(b) [(d)]. When we employ a much smaller value of $U$ of 0.4 eV, 
which is close to the value suggested by Fig. 3(c) in Ref.~\onlinecite{Xue_prl_2DK3C60_SC}, a sharp metal-insulator transition is absent, as shown in Fig.~\ref{fig:U0px_Jm0p03_muS_0p7}(a).

\begin{figure*}[htp]
\centerline{\epsfig{figure=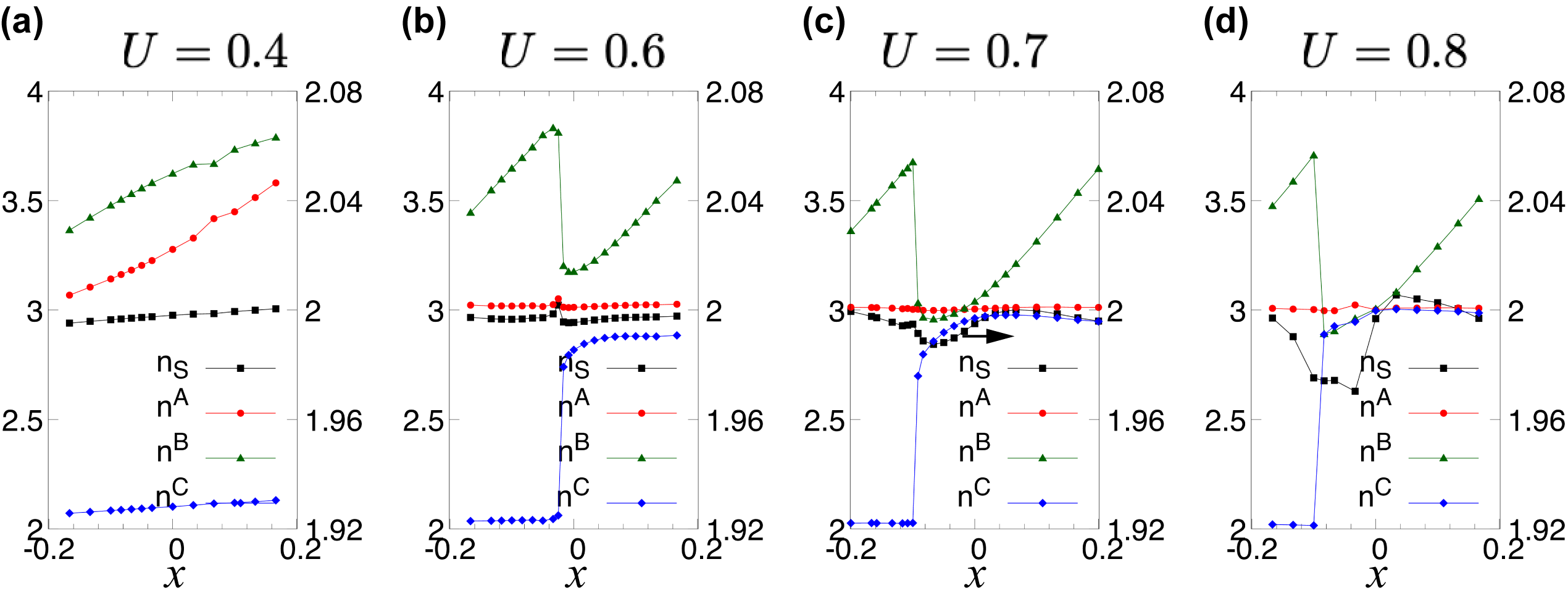,width=0.8\columnwidth}}
\caption{ 
Fillings in the different layers and in the substrate for the indicated interactions $U$
at $T=50$ K, $\mu_S=0.7$ eV, $J=-0.03$ eV. 
Panel (a) $U=0.4$ eV.
Panel (b) $U=0.6$ eV.
Panel (c) $U=0.7$ eV.
Panel (d) $U=0.8$ eV.
}
\label{fig:U0px_Jm0p03_muS_0p7}
\end{figure*}

\subsection*{SM4. Determination of the substrate energy shift $\mu_S$ and its effects on the results}

${\mathcal H}({\bf k})$ 
 describes 3 narrow $t_{1u}$ bands for each of the 3 fulleride layers, plus 3 dispersive bands for 
the substrate. This setup resembles the ``d+p" model which is often used in the description of transitional metal oxides \cite{Dang2014}. In our case, the $t_{1u}$-bands
play the role of correlated $d$-bands, while the dispersive bands of the substrate play the role of weakly-interacting $p$-bands.

The additional shift $\mu_S$ to the substrate in Eq.~(\ref{eq:DMFT_eq_1}) is necessary to reproduce the experimentally observed $n=3$ Mott insulating state of the monolayer system. Here, we note that the almost particle-hole symmetric STM signal of the monolayer system \cite{Xue_prl_2DK3C60_SC} clearly shows that this insulator is $n=3$ and not $n=2$. 
When $U=0.7$ eV, we find that the monolayer is a $n=3$ Mott insulator for $0.55<\mu_S<0.85$, as shown in Fig.~\ref{fig:hartree_shift_U0p7}. If $U$ is smaller (bigger), the range of $\mu_S$ where the monolayer is Mott insulating becomes narrower (wider), but always with values around 0.7 eV, as shown in 
Fig.~\ref{fig:hartree_shift_U0p6} and Fig.~\ref{fig:hartree_shift_U0p8} for 
$U=0.6$ and $U=0.8$ eV, respectively. As a result, we choose $\mu_S=0.7$ in the DFT+DMFT calculation for both the monolayer and trilayer system. 

\begin{figure*}[h!]
\centerline{\epsfig{figure=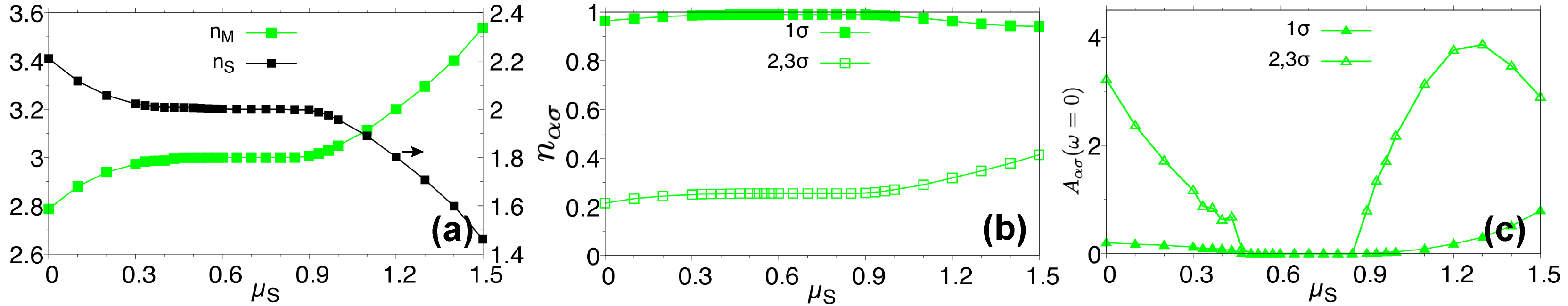,width=0.7\columnwidth}}
\caption{ 
Determination of the extra Hartree energy shift $\mu_S$ of the substrate for the monolayer+substrate system with $U = 0.7$ eV, $J=-0.03$ eV ($T=50$ K) and total filling fixed at the nominal $n=5$.
Panel (a): fillings in the monolayer ($t_{1u}$ orbitals, green) and the substrate (black) as a 
function of the shift $\mu_{S}$. Panel (b): spin-orbital resolved fillings in the monolayer as a function of $\mu_S$. 
Panel (c): spin-orbital resolved zero-frequency spectra in the monolayer as a function of $\mu_S$. 
}
\label{fig:hartree_shift_U0p7}
\end{figure*}

\begin{figure*}[h!]
\centerline{\epsfig{figure=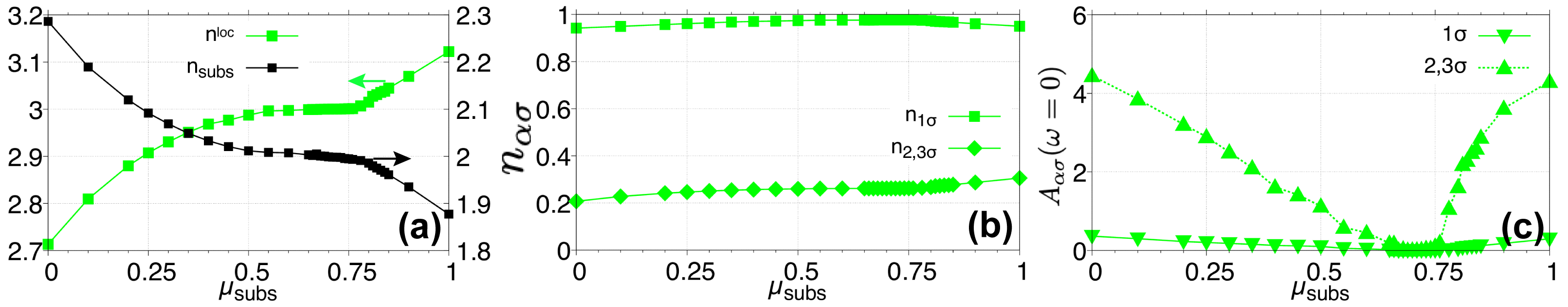,width=0.7\columnwidth}}
\caption{Similar to Figure.~\ref{fig:hartree_shift_U0p7}, but here $U=0.6$ eV.
}
\label{fig:hartree_shift_U0p6}
\end{figure*}

\begin{figure*}[h!]
\centerline{\epsfig{figure=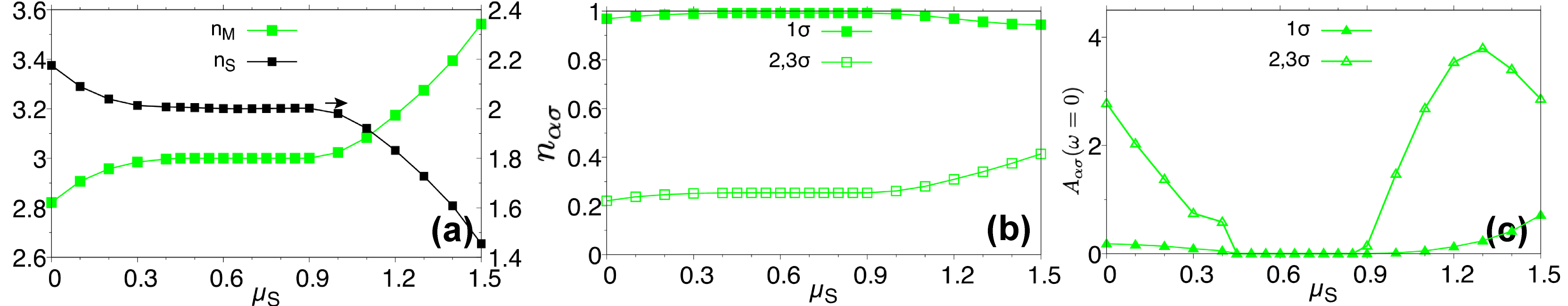,width=0.7\columnwidth}}
\caption{Similar to Figure.~\ref{fig:hartree_shift_U0p7}, but here $U=0.8$ eV.
}
\label{fig:hartree_shift_U0p8}
\end{figure*}

Last but not least, we would like to mention that our results are essentially independent of $\mu_S$, as long as $\mu_S$ lies inside the relevant range of $\mu_S$. As shown in Figs.~\ref{fig:U0p7Jm0p03_muS_x}(b)-(d), the results with $\mu_S$ = 0.6--0.8 eV all capture the electron-hole asymmetry observed in the experiments. On the other hand, the Mott transition point occurs on the electron doped side without a shift [$\mu_S$ = 0.0, Fig.~\ref{fig:U0p7Jm0p03_muS_x}(a)], which is not consistent with the experiments, where the transition appears on the hole doped side. Hence, although a relative shift between the weakly correlated and strongly correlated bands is necessary, as in other LDA+DMFT studies, our results are robust against modifications in $\mu_S$ on the order of 0.1 eV, and our statements do not rely on any particular fine-tuning. (Note that the width of the bands is only about 0.4 eV, so a shift in a range of 0.1--0.2 eV is significant.)

\begin{figure*}[htp]
\centerline{\epsfig{figure=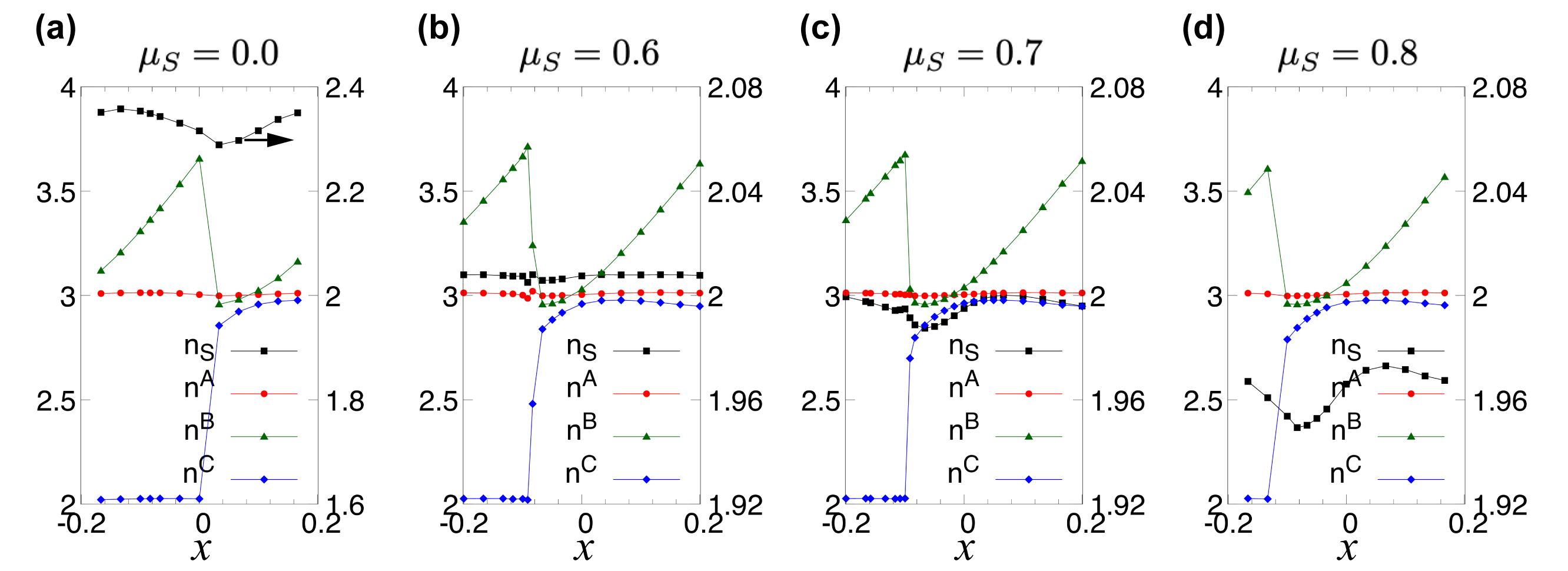,width=0.7\columnwidth}}
\caption{
Fillings in the different layers and in the substrate for different shifts $\mu_S$
at $T=50$ K, $U=0.7$ eV, $J=-0.03$ eV. 
Panel (a) $\mu_S=0.0$ eV.
Panel (b) $\mu_S=0.6$ eV.
Panel (c) $\mu_S=0.7$ eV.
Panel (d) $\mu_S=0.8$ eV.
}
\label{fig:U0p7Jm0p03_muS_x}
\end{figure*}

\subsection*{SM5. Results for a rotationally invariant interaction}

The realistic local interaction in K$_3$C$_{60}$ is of the Slater-Kanamori type, with pair-hopping $J_p$ and spin-flip $J_s$ terms. In our study, we ignore the pair-hopping and spin-flip terms, because of the computational cost associated with treating these terms and the fact that the $|J|$ value is small. 
To estimate the effect of the neglected terms, we have performed two computationally expensive calculations for $x=-0.1$, in the Mott phase, and for $x=0$, in the metallic phase, using $U=0.7$ eV and $J_s=J_p=J=-0.03$ eV at $T=10$ K. 

Figure~\ref{fig:Aw_pairhoping_U0p7Jm0p03_T10K} shows the top layer normal-state spectral functions using the Slater-Kanamori type interactions.
We do observe the metal-insulator transition in the top layer on the hole doped side, between $x=-0.1$ and $x=0$. 
As shown in Tab.~\ref{tab:RotInv_U0p7_T10K}, at the metal-insulator transition, the filling of the top layer is drastically reduced to $n=2$, as a result of charge transfer between the top and middle layers. 
Therefore, the scenario for the electron-hole asymmetry discussed in the main text holds even in the case of the Slater-Kanamori type interaction. 

\begin{figure*}[htp]
\centerline{\epsfig{figure=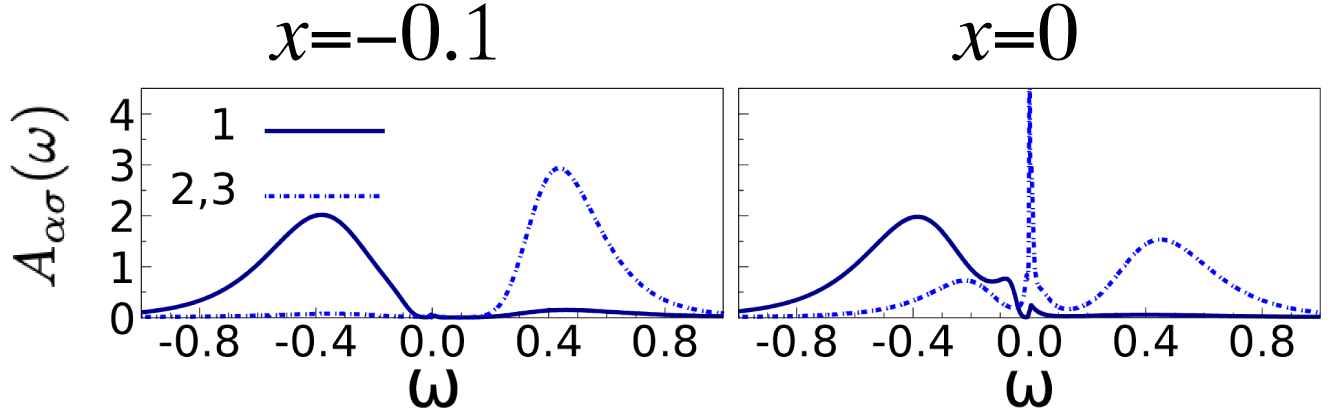,width=0.6\columnwidth}}
\caption{ 
Orbital-resolved local spectra $A_{\alpha\sigma}(\omega)$ in the top layer 
obtained by DFT+DMFT for the trilayer+substrate system with rotationally invariant interactions. 
The thick lines correspond to orbital $1$ and the dashed lines to orbitals $2,3$.
The parameters are  $U=0.7$ eV, $J_s=J_p=J=-0.03$ eV, $T$=10 K.
}
\label{fig:Aw_pairhoping_U0p7Jm0p03_T10K}
\end{figure*}
\begin{table*}[!h]
\centering
\caption{Fillings per layer obtained by DFT+DMFT for the trilayer+substrate system with the rotationally invariant interactions.
The parameters are  $U=0.7$ eV, $J_s=J_p=J=-0.03$ eV, $T$=10 K.}
\label{tab:RotInv_U0p7_T10K}
\medskip
\begin{tabular}{ccccc}
\hline
Doping &  $n_A$  & $n_B$ & $n_C$ & $n_S$ \tabularnewline
\hline
$x=0$ & 3.03 & 2.98 & 2.98 & 2.02  \tabularnewline
\hline
$x=-0.1$ & 3.01 & 3.66 & 2.03 & 1.99 \tabularnewline
\hline
\end{tabular}
\end{table*}

\end{flushleft}
\end{widetext}

\end{document}